\documentclass[twocolumn,showpacs,aps,prl,superscriptaddress]{revtex4}

\usepackage{graphicx}
\usepackage{dcolumn}
\usepackage{amsmath}
\usepackage{epsfig}
\input pubboard/babarsym
\newcommand{\BABARPubYear}    {04}
\newcommand{\BABARPubNumber}  {45}

\newcommand{\SLACPubNumber} {10938}
\newcommand{\LANLNumber} {0501038}

\def\figurebox#1#2#3{
    \def\arg{#3}
    \ifx\arg\empty
    {\hfill\vbox{\hsize#2\hrule\hbox to #2{\vrule\hfill\vbox to #1{\hsize#2\vfill}\vrule}\hrule}\hfill}
    \else
    {\hfill\epsfbox{#3}\hfill}
    \fi}

\begin{document}

\preprint{\babar-PUB-\BABARPubYear/\BABARPubNumber} 
\preprint{SLAC-PUB-\SLACPubNumber} 

\begin{flushleft}
\babar-PUB-\BABARPubYear/\BABARPubNumber\\
SLAC-PUB-\SLACPubNumber\\
hep-ex/\LANLNumber\\[10mm]
\end{flushleft}

\title{
{\large \bf
Search for the Radiative Decay $\Bz\rightarrow\phi\gamma$} 
}

%% author list as of 02-Nov-2004 (611 authors)
%
\author{B.~Aubert}
\author{R.~Barate}
\author{D.~Boutigny}
\author{F.~Couderc}
\author{Y.~Karyotakis}
\author{J.~P.~Lees}
\author{V.~Poireau}
\author{V.~Tisserand}
\author{A.~Zghiche}
\affiliation{Laboratoire de Physique des Particules, F-74941 Annecy-le-Vieux, France }
\author{E.~Grauges-Pous}
\affiliation{Universitad Autonoma de Barcelona, E-08193 Bellaterra, Barcelona, Spain }
% ...
\collaboration{The \babar\ Collaboration}
\noaffiliation

\date{\today}

\begin{abstract}
We perform a search for the exclusive radiative decay $\Bz\rightarrow\phi\gamma$, which is dominated by $\overline b d$ annihilation, in a sample of 124 million $\BB$ events collected with the \babar\ detector 
at the \pep2\ asymmetric-energy $\epem$ storage ring at SLAC. No significant signal is 
seen. We set an upper limit on the branching fraction of 
$\BR(\Bz\rightarrow \phi\gamma)<8.5\times 10^{-7}$ at the 90\% confidence level. 
\end{abstract}


\maketitle
\begin{figure}[t]
\epsfig{file=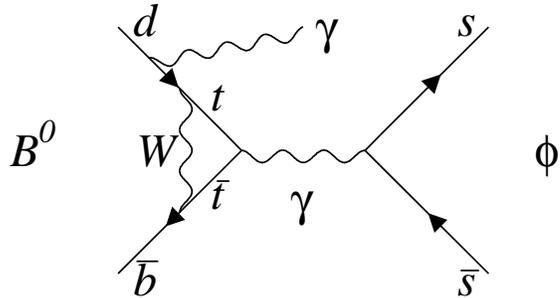,width=0.9\columnwidth}
\caption{One of the leading order Feynman diagrams contributing to the decay 
$\Bz\rightarrow\phi\gamma$ in the Standard Model.}
\label{fig:btophig-feyn}
\end{figure}

Within the Standard Model (SM) the rare decay 
$\Bz\rightarrow\phi\gamma$ proceeds through a penguin annihilation 
process.  No process of this kind has yet been observed.  The largest short-distance contribution to the SM amplitude is illustrated in 
Fig.~\ref{fig:btophig-feyn}. The coupling of the top quark within the 
loop to the $\d$ quark leads to a dependence of the amplitude on the 
Cabbibo--Kobayashi--Maskawa matrix element $V_{td}$ \cite{CKM} that suppresses the 
decay rate.  Helicity suppression and the smallness of the decay constants $f_{B}$ and $f_{\phi}$ also lead to a small decay rate~\cite{LLWY}.  A recent QCD factorization 
calculation predicts a SM branching fraction of $3.6\times10^{-12}$ \cite{LLWY}. However, contributions to the 
$\Bz\rightarrow\phi\gamma$ amplitude are possible from physics beyond the SM 
where new heavy particles enter the loop. For example, some models of 
supersymmetry that violate $R$-parity predict an enhancement of 
the $\Bz\rightarrow\phi\gamma$ branching fraction by up to 
four orders of magnitude \cite{LLWY}. A prior experiment has bounded 
the branching fraction to be $\BR(\Bz\rightarrow \phi\gamma)<3.3\times 10^{-6}$ at 
the 90\% confidence level (C.L.) \cite{CLEO-phig}. 

This analysis uses data collected with the $\babar$ detector at the \pep2\ 
asymmetric-energy $\epem$ storage ring at SLAC. 
The data sample consists of $124.1\pm1.4$ million $\BB$ events, corresponding to an 
integrated luminosity of $113~\invfb$ on the $\FourS$ resonance, which has a mass of 10.58 $\gevcc$. 

The $\babar$ detector is described in detail in Ref. \cite{BABARNIM}.  Charged particle trajectories are measured by the combination of a 
five-layer silicon vertex tracker (SVT) and a 40-layer drift chamber (DCH), which 
are embedded in the $1.5~\mathrm{T}$ magnetic field of a solenoid. Photons are 
detected in a CsI(Tl) crystal electromagnetic calorimeter (EMC), with an energy 
resolution of $\sigma_{E}/E = 0.023(E/\gev)^{-1/4}\oplus 0.019$.  Charged hadron identification 
is performed by the combination of energy-loss $(\dedx)$ information from the SVT and DCH and measurements from a ring-imaging Cherenkov detector (DIRC).  The segmented flux return (IFR) of the magnet is instrumented with resistive plate chambers to identify muons.

Monte Carlo (MC) simulations of the response of the $\babar$ detector, based on GEANT4 
\cite{geant4}, are used to optimize the selection criteria and determine 
the signal efficiencies. These simulations take into account the variations  
of detector conditions and beam induced backgrounds during the data--taking 
period. 

The first stage of the analysis is to identify $\phi$ mesons and high-energy photons, 
which can be combined to form $\Bz$ candidates \cite{CC}.  The $\phi$ is reconstructed 
in the decay to $\Kp\Km$, which corresponds to $(49.2^{+0.6}_{-0.7})\%$ of the total 
$\phi$ decay rate \cite{PDG2002}. We require that the charged tracks used in reconstructing the $\phi$ 
candidates have associated hits in both the SVT and DCH and have a 
transverse momentum greater than $0.1 \gevc$.  Tracks compatible with being kaons are identified via an algorithm that combines the candidate 
track's measured $\dedx$, Cherenkov angle and number of Cherenkov photons.  Individual kaons produced by 
signal $\phi$ meson decays are identified with 80\% efficiency by this algorithm, 
while the misidentification rate of pions as kaons is less than 2\% over most of 
the relevant kaon momentum range. A $\phi$ candidate is composed of two identified 
kaons of opposite charge 
that are consistent with originating from a common vertex. We require  
$\phi$ candidates have a mass of  $1.011<m_{\Kp\Km}<1.029~\gevcc$ (a full width 4.2 times the natural width); this criterion is optimized 
by a procedure described below.

A photon is identified as a shower in the EMC that is not associated with a reconstructed track.   We remove poorly reconstructed photons by rejecting showers in crystals with high noise rates or rejecting photons that are near the edge of the calorimeter's acceptance.  Furthermore, showers that contain energy 
deposits measured by crystals adjacent to an inefficient crystal are rejected. 
The shower profile is required to be compatible with a single photon to reject those that arise from $\piz$ decays where the two photons
enter the calorimeter in close proximity to one another; this condition also rejects 
showers generated by neutral hadrons, predominantly $\overline{n}$ and $\KL$, that
have a broader lateral profile than photon showers. To suppress photons from 
$\piz$ $(\eta)$ decays, the photon candidate is combined in turn with all other photons 
in the event with a laboratory energy greater than $50~(250)\mev$.  If any of the resulting invariant $\gamma\gamma$ masses are within 
$20~(40)\mevcc$ of the $\piz$ $(\eta)$ mass, the candidate is vetoed. 
To further remove photons from $\piz$ decays the shower is required to be 
isolated by at least $25~\cm$ from any other shower in the event.
\begin{figure}[t]
\epsfig{file=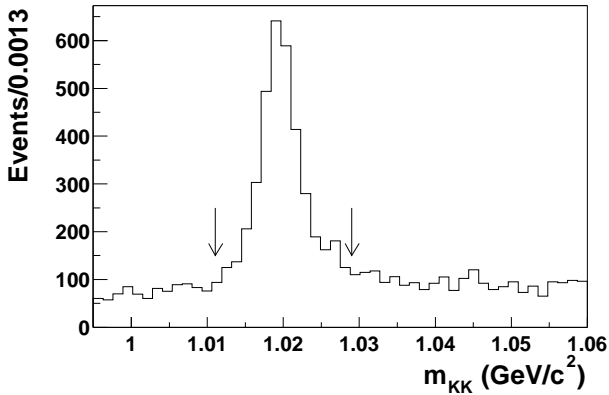,width=1.0\columnwidth}
\caption{The $\phi$ invariant mass distribution of on-resonance data.  The $\phi$ candidates are selected in the region indicated by the arrows.  All selections have been applied except the $\mes$, $\DeltaE^{*}$, $\phi$ mass, and neural net requirements. The distribution is dominated by combinatorial selections.}
\label{Massonpeak}
\end{figure}

The photon and $\phi$ meson candidates are combined to form $\Bz$ meson 
candidates. We define $\DeltaE^{*} \equiv E^{*}_{B}-E^{*}_{\mathrm{beam}}$, where 
$E^{*}_{B}$=$E^{*}_{\phi}+ E^{*}_{\gamma}$ is the center-of-mass (CM) energy of the 
$\Bz$ meson candidate and $E^{*}_{\mathrm{beam}}$ is the CM energy of each 
beam. The signal $\DeltaE^{*}$ distribution is peaked at zero 
with a resolution of approximately $50\mev$; there is a negative 
tail in $\DeltaE^{*}$ because of the asymmetric $E^{*}_{\gamma}$ resolution 
that is the result of energy leakage from the EMC. We also define the 
beam-energy-substituted mass 
$\mes \equiv \sqrt{E^{*2}_{\mathrm{beam}}-\mathbf{p}^{*2}_{B}}$, where 
$\mathbf{p}^{*}_{\B}$ is the CM three-momentum of the $\Bz$ candidate.
The signal $\mes$ distribution peaks at the mass of the $B$ meson,
$m_B=5.279~\gevcc$ \cite{PDG2002}, and has a resolution of $3\mevcc$, which
is dominated by the spread in $E^{*}_{\mathrm{beam}}$.   The signal region in which we search for 
$\Bz\rightarrow\phi\gamma$ events is defined as 
$-0.2<\DeltaE^{*}<0.1~\gev$ and $5.27<\mes<5.29~\gevcc$. We define three sideband regions for estimating backgrounds: (I) $5.10 < \mes < 5.29~\gevcc$ and $0.1<\DeltaE^{*}<0.5~\gev$, (II) $5.10 < \mes < 5.29~\gevcc$ and $-0.5<\DeltaE^{*}<-0.2~\gev$, and  (III) $5.10 < \mes < 5.27~\gevcc$ and $-0.2<\DeltaE^{*}<0.1~\gev$.
At this stage in the analysis all candidates with $-0.5<\DeltaE^{*}<0.5~\gev$ and $5.10<\mes<5.29~\gevcc$
are retained.

\begin{figure}[t]
\epsfig{file=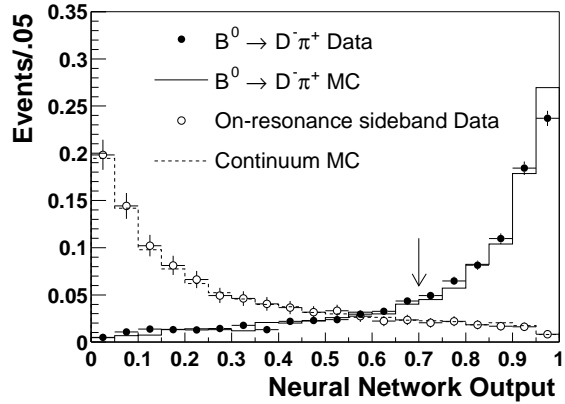,width=1.0\columnwidth}
\caption{Neural network output for MC-simulated events with comparison to data control samples.  The histograms have been normalized such that their total area is unity.  The on-resonance sideband is defined as regions (I) and (II) in the text.  The arrow indicates the lower limit imposed on the data. }
\label{NNout}
\end{figure}

The background comes predominantly from random combinations of 
real $\phi$ mesons (Fig.~\ref{Massonpeak}) and high-energy photons produced in continuum $\u$, $\d$, $\s$ and 
$c$ quark-antiquark events.  The dominant sources of high-energy photons in 
continuum events are initial-state radiation (ISR) and 
$\piz/\eta\rightarrow\gamma\gamma$ decays where the second photon 
is undetected, or the measured two-photon mass lies outside the $\piz/\eta$ veto window.  Since the continuum background does not peak in $\DeltaE^{*}$ or $\mes$, its 
magnitude can be evaluated from parts of the $\mes$ and $\DeltaE^{*}$ data distributions in the sideband regions. There are also potential backgrounds from 
charmless \B decays that peak in $\mes$ and $\DeltaE^{*}$. The \BB background is estimated from simulation.    

The ratio of the second--to--zeroth Fox--Wolfram moments \cite{fox}, for charged tracks in the event measured in the CM frame, is required to be less than 0.9 to reject some low multiplicity continuum final 
states. The combinatorial background within the signal region is reduced further 
by combining 24 input variables that distinguish between $\qqbar$ continuum
and $\BB$ events into one discriminating variable via a neural network, as in 
Refs.~\cite{prlkstgam,prlrhogam}. The neural network responds non--linearly 
to the input variables and exploits correlations among them \cite{SNNS}.  To discriminate between 
jet-like continuum background and the more spherically-symmetric signal events, we include in the neural network the absolute value of the cosine of the angle between the high-energy photon and the CM thrust axis of 
the reconstructed particles in the r.o.e (rest of the event), $|\cos{\theta^{*}_{T}}|$, and the energy distribution of all reconstructed particles in the r.o.e.~binned into eighteen $10\degrees$ intervals around the photon direction. 
The distribution of events in $|\cos{\theta^{*}_{T}}|$ is uniform for signal events but is strongly peaked toward 1 for continuum events. The energy flow is collinear with the photon direction in continuum 
events but is more isotropic in signal events. To discriminate against ISR continuum events we incorporate in the neural network the ratio of the second--to--zeroth Fox-Wolfram 
moments computed in the photon recoil frame, in which ISR continuum events 
are more likely than signal events to have a back--to--back jet structure.  
To provide further discrimination from continuum background, we include in the selection two angular variables of the signal decay: $|\cos{\theta_{B}^{*}}|$, where $\theta_{B}^{*}$
is the angle between the $B$ meson candidate's momentum and the beam axis in the CM frame, and $|\cos{\theta_H}|$, where $\theta_H$ is the angle between the flight 
directions of one of the daughters of the $\phi$ meson and the $B$ meson 
candidate in the rest frame of the $\phi$ meson. The distribution of events of both of these variables is proportional to $\sin^2{\theta}$, where $\theta$ is $\theta_{B}^{*}$ or $\theta_H$ for signal events. The $|\cos{\theta_{B}^{*}}|$ distribution has this form because a vector state (\FourS) is decaying to two pseudoscalars ($\BB$), with the vector state having no helicity zero component.  The $|\cos{\theta_H}|$ distribution has this form because a pseudoscalar state $(\Bz)$ is decaying to two vector states ($\phi\gamma$), with 
the photon having no helicity zero component.     
Finally, to provide further sensitivity to $B$ decays in the event, we admit two more variables.  The first is the longitudinal separation between the decay vertex of the 
$B$ meson candidate and the vertex of the other charged particles in the 
event, which tends to be non--zero for signal events due to the long 
lifetime of the $B$ meson, whereas all particles in continuum events usually  
originate from a single vertex.  The second is the net flavor of the rest of the event, $N_{\mathcal{F}}=N_{\KS}+\sum_{i}|N_{i}^{+}-N_{i}^{-}|$, 
where $N_{\KS}$ is the number of reconstructed $\KS$ and $N_{i}^{\pm}$ 
are the numbers of reconstructed charged particles of type $i=\epm,\mu^{\pm},\Kpm~\mathrm{or}~\pipm_{\mathrm{slow}}$ \cite{slowpi}.  The discrimination power of this variable can be seen by noting how the final state particles (as seen by the detector) are produced.  If the final state particle is produced through the decay of a $B$ meson, then it most likely was generated through the weak interaction, which allows for flavor-changing currents.  In contrast, if it is part of continuum $\u$, $\d$, and $\s$ quark-antiquark events, then there is no net flavor production.

The neural network is trained on samples of simulated signal and continuum events.  A $B\rightarrow D\pi$ control sample was used to determine the systematic error in the selection efficiency of the neural network.  Here, the ``bachelor pion'' is used in place of the signal photon.  This sample is representative of signal events because the input variables are mainly derived from particles not associated with the $B$ candidate.  The exception is $|\cos{\theta_H}|$, which has a distribution for pions different from that for photons because of the different spins of these particles..  Instead, this variable is drawn from the signal distribution of $\sin^2{\theta}$.  Fig.~\ref{NNout} shows the distribution of the neural network output for MC-simulated $B^{0} \rightarrow D^{-}\pi^{+}$ events compared to $B^{0} \rightarrow D^{-}\pi^{+}$ events reconstructed in on-resonance data. Also shown is the distribution of the neural network output for MC simulated continuum background events compared to events in the on-resonance sideband. These distributions show reasonable agreement between MC and data samples. 

The selection is  optimized to achieve the best upper limit because of the very 
small SM expectation for $\BR(\Bz\rightarrow\phi\gamma)$. The quantity minimized by the 
optimization is $N/\epsilon$, where $N$ is the average 90\% confidence level upper limit on the 
number of reconstructed signal events and $\epsilon$ is the signal efficiency.  $N$ is estimated from an ensemble of experiments with a given MC-simulated background and no true 
signal \cite{FAndC}. 
We optimize the selection on the neural network output and on the $\phi$ mass window simultaneously. The photon selection is identical to that optimized for the measurement of $\B\to\Kstar\gamma$ \cite{prlkstgam}.  
The continuum background within the signal region is estimated with likelihood 
fits to the $\DeltaE^{*}$ and $\mes$ distributions of simulated continuum events 
for each set of selection criteria considered during the optimization. The probability density functions 
(PDFs) used to describe the $\mes$ and $\DeltaE^{*}$ distributions 
are an ARGUS threshold function \cite{argus} and a first-order polynomial, respectively. The resulting 
functions are integrated over the signal region to estimate the background. 
The optimized selection criteria have a signal selection efficiency of $(14.4\pm0.1)\%$ and a mean
continuum background yield from MC simulation of $4.4\pm0.5$ events.  

The selection leads to an expectation of $0.073\pm0.004$ \BB background events within the signal region, as determined from the MC simulation. 
The decays $\Bz\rightarrow\phi\piz$ and $\Bz\rightarrow\phi\eta$ contribute most of the events; 
these charmless modes have not been observed, so the largest branching fractions predicted within the SM are
used to compute the expectations \cite{charmless}. The remaining contributions come from $\Bz\rightarrow\phi\eta^{\prime}$,
$\Bz\rightarrow\phi\KS$ and $\Bz\rightarrow\phi\KL$ decays. The contributions of all other \BB events, including those with $\Bz\rightarrow\Kstarz\gamma$, were found to be negligible.

To make a more robust estimate of the continuum background we use the data themselves; this eliminates any uncertainty related to 
the simulation of the \qqbar continuum.  To define functions that describe the background distribution we use a maximum-likelihood fit to the data within sideband regions I and II with the same form of the $\mes$ and $\DeltaE^{*}$ PDFs as those used in the optimization
procedure. The resulting functions are integrated over the sideband region III   
and the signal region to estimate the amount of continuum background within 
them. To validate the method 
the number of background expected in region III (mean of $79\pm6$ events) is compared to the number of events observed in this 
interval ($71$ events); these two numbers are in good agreement.
To assign the error on the background estimate, we take the difference of both PDFs evaluated with the central value of the fit parameters and the parameters varied by one standard deviation and then add the differences in quadrature.  The mean continuum background estimate within the signal region is $6.0\pm0.9$ events.

In addition, we perform other consistency tests to ensure that both the background estimate and the method used to acquire it are reasonable.  We use 
a second-order polynomial as the PDF for $\DeltaE^{*}$; this leads to a mean continuum
background estimate of $5.6\pm0.8$ events. The difference between 5.6 and 6.0 (the number obtained with the method defined previously) is taken as a systematic uncertainty on the 
background. 
To check the assumption that the PDFs factorize, we compute the ARGUS 
parameter and the slope of the first-order polynomial in different intervals of $\DeltaE^{*}$ and \mes; the parameters are found to be in reasonable agreement.  
We perform a fit to \mes and $\DeltaE^{*}$ using regions I, II and III; this
gives a mean expected background of $5.2\pm0.8$ events, which is in agreement with the expectation from using only regions I and III.  The mean continuum background estimate within the signal region including the systematic uncertainty is $6.0\pm1.0$ events.

\begin{figure}[t!]
\epsfig{file=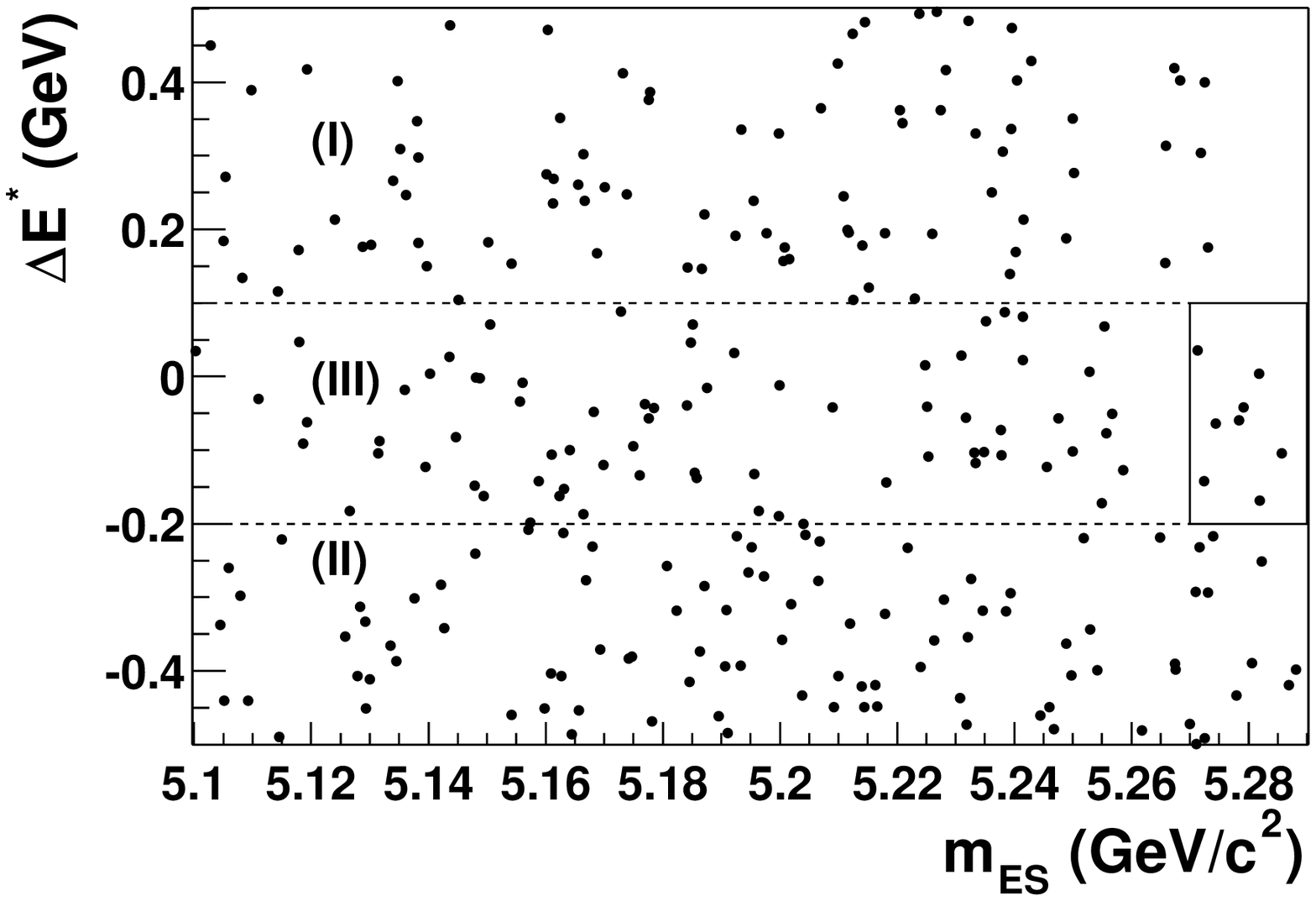,width=1.0\columnwidth}
(a)
\epsfig{file=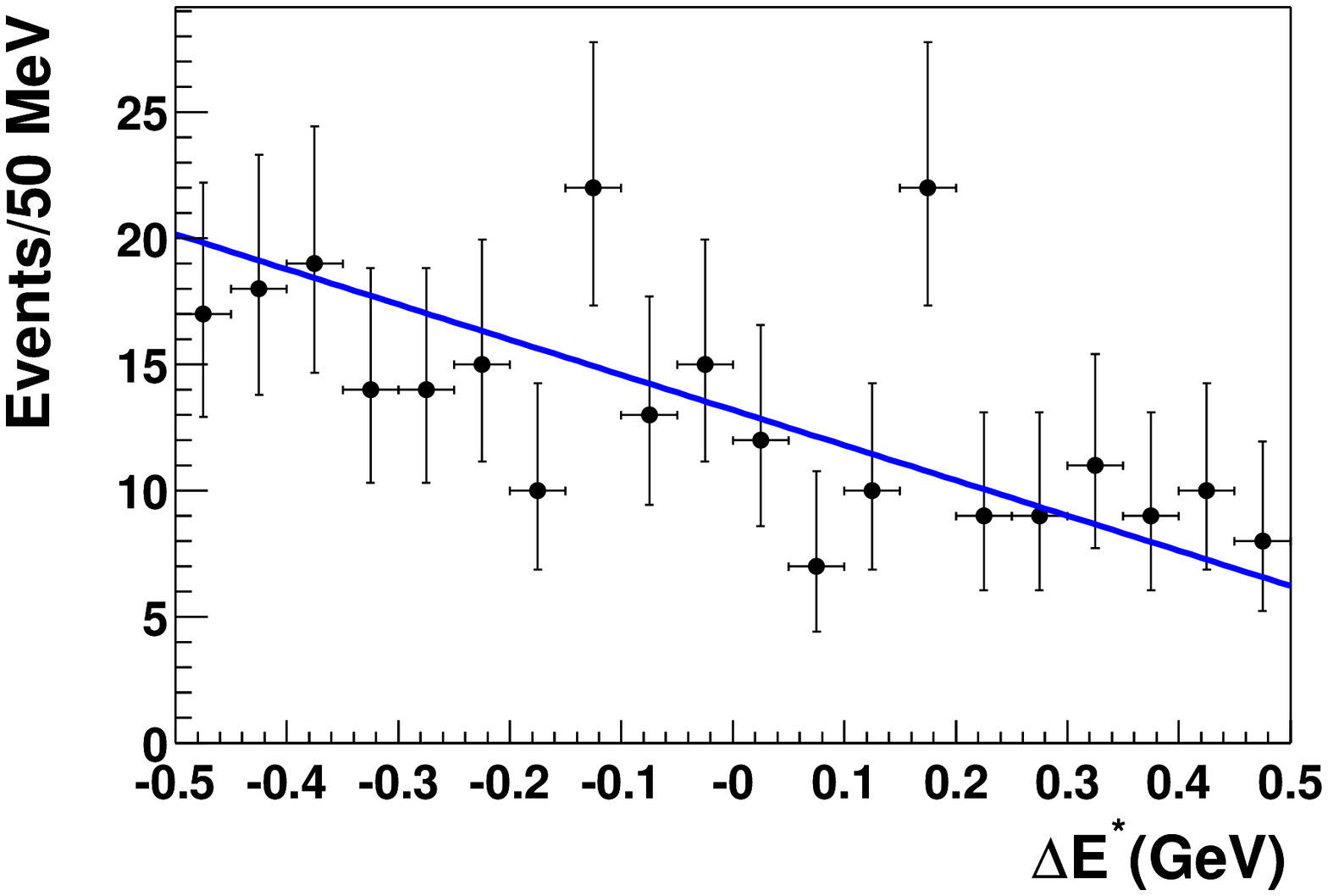,width=1.0\columnwidth}
(b)
\epsfig{file=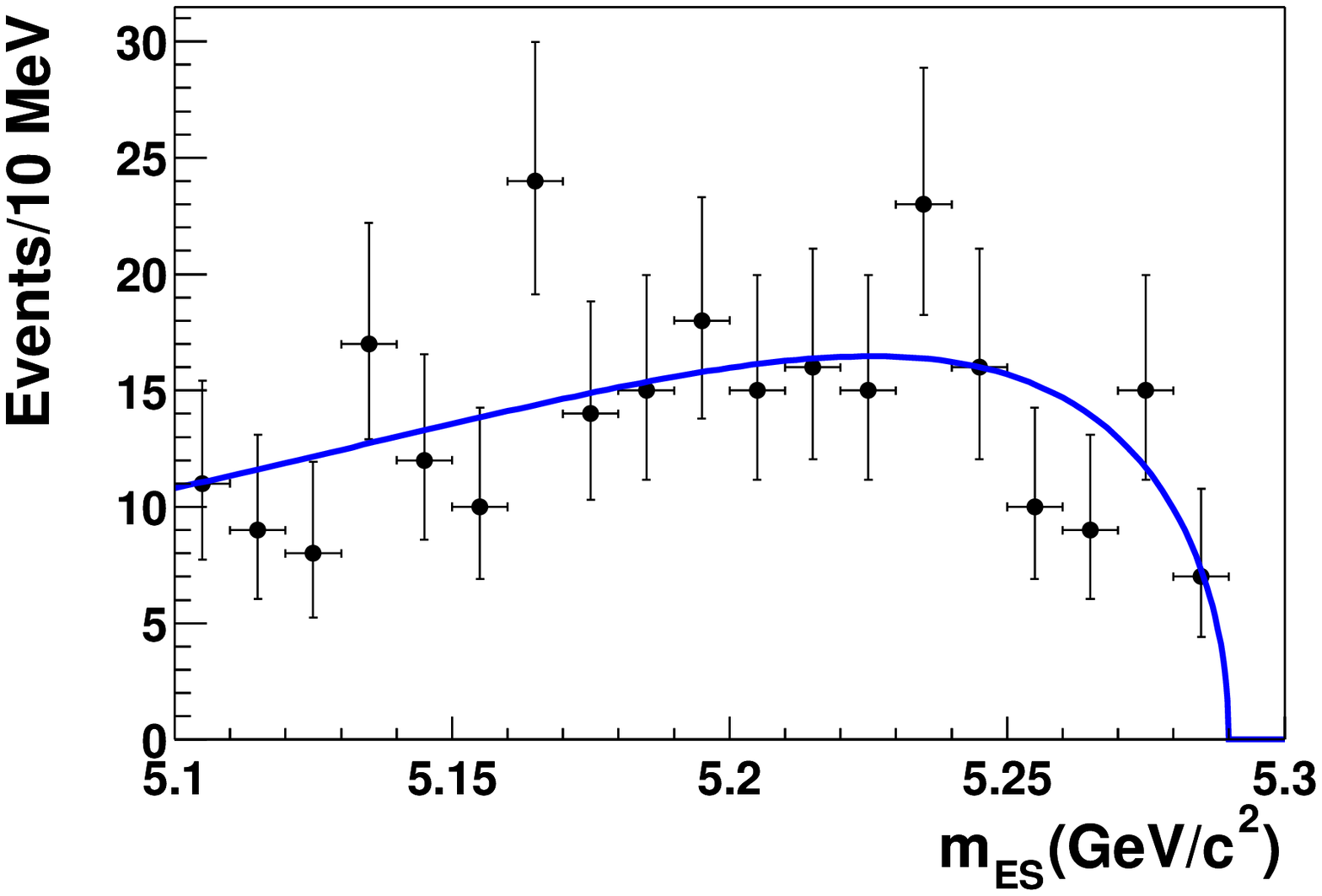,width=1.0\columnwidth}
(c)
\caption{In (a), the $\mes$ - $\DeltaE^{*}$ distribution of data events after all selection criteria have been applied is shown. The box with the solid line indicates the signal region, while the dashed lines indicate the regions defined in the text. In (b) and (c) are  the $\DeltaE^{*}$ and $\mes$ projections respectively of (a).}
\label{dataMesVsDeltaE}
\end{figure}

We observe 8 events in the signal region, which is consistent with the background 
estimation. The distribution of data events in $\mes$ and $\DeltaE^{*}$, together with the projections in these two variables, can be seen in Fig. \ref{dataMesVsDeltaE}. 

The fractional systematic uncertainties on $\varepsilon$, the number of $\BzBzb$ pairs in the data set ($N_{\BzBzb}$), and $\BR(\phi\rightarrow\Kp\Km)$ are 
summarized in Table~\ref{effsyst}; their origins are briefly described below.
The difference in 
tracking efficiency between data and simulation is evaluated with a 
sample of tracks that are well reconstructed in the SVT; these studies yield a relative 
shift of $(-0.8\pm 1.3)\%$ per charged track. The uncertainty related to the kaon identification is evaluated with
simulation and data samples of the decay $D^{*+}\rightarrow\Dz(\rightarrow\Km\pip)\pip$ and its charge conjugate. The photon identification 
efficiency is studied in samples of 
$\piz$ decays from $\tautau$ events.  The uncertainties and corrections due to the $\piz$ and $\eta$ vetoes and shower 
isolation criteria are evaluated by embedding simulated photons into both data and simulation samples of \BB events.   The difference between 
the neural network selection efficiency in the $B^{0} \rightarrow D^{-}\pi^{+}$ data and simulation samples is $(-1.5\pm 2.7)\%$. There is 
a small uncertainty related to the statistics of the signal simulation sample used to calculate the efficiency. The corrected value of the efficiency after all corrections are applied is $(13.9\pm 0.7)\%$. 
The number of \BB events used in the analysis is calculated from the change in the ratio of multihadron to $\mumu$
events between data taken at the $\FourS$ resonance and at $\epem$ CM energy $40\mev$ below the resonance; there is 
$1.1\%$ fractional uncertainty in this measurement.  We assume that half the number of \BB events are \BzBzb events.  The measured 
branching fraction $\BR(\phi\rightarrow\Kp\Km)$ has 
a fractional uncertainty of 0.6\% \cite{PDG2002}.       

\begin{table}[t]
\begin{tabular}{lcc}\hline\hline
Systematic effect            & Correction & Uncertainty (\%) \\ \hline
$\Kp\Km$ tracking            &   0.984    &  2.6 \\
$\Kp\Km$ identification      &     -      &  2.0 \\
Shower separation            &     -      &  2.0 \\
$\piz/\eta$ veto             &     -      &  1.0 \\
Photon detection efficiency  &   0.997    &  2.5 \\
Continuum suppression        &   0.985    &  2.7 \\
Simulation statistics        &     -      &  0.1 \\ 
\hline
Overall signal efficiency  &   0.966    &  5.4 \\
\hline
\BB counting                 &     -      &  1.1 \\
$\BR(\phi\rightarrow\Kp\Km)$ &     -      &  0.6 \\ \hline\hline 
\end{tabular}
\caption{The sources of systematic uncertainty on estimating the branching fraction. Any correction 
factor applied to the signal efficiency related to the systematic effect is also given.}\label{effsyst} 
\end{table}

Using the signal efficiency, $\BR(\phi\rightarrow\Kp\Km)$, $N_{\BzBzb}$, the background estimation along with the associated uncertainties, we find by the procedure of Ref. \cite{Barlow} the upper limit: 
\begin{displaymath}
	 \BR(\Bz\rightarrow\phi\gamma)<8.5\times 10^{-7}, 
\end{displaymath}
at the 90\% C.L.  In conclusion, no evidence for the decay $\Bz\rightarrow\phi\gamma$ is observed. We set an upper limit that 
is 3.9 times lower than the previously published result.  
      
We are grateful for the excellent luminosity and machine conditions
provided by our \pep2\ colleagues, 
and for the substantial dedicated effort from
the computing organizations that support \babar.
The collaborating institutions wish to thank 
SLAC for its support and kind hospitality. 
This work is supported by
DOE
and NSF (USA),
NSERC (Canada),
IHEP (China),
CEA and
CNRS-IN2P3
(France),
BMBF and DFG
(Germany),
INFN (Italy),
FOM (The Netherlands),
NFR (Norway),
MIST (Russia), and
PPARC (United Kingdom). 
Individuals have received support from CONACyT (Mexico), A.~P.~Sloan Foundation, 
Research Corporation,
and Alexander von Humboldt Foundation.


\begin{thebibliography}{99}
\bibitem{CC}
Charge conjugate states are implied throughout this paper.
\bibitem{slowpi}
A $\pi^{\pm}_{\mathrm{slow}}$ is defined as a pion having a CM momentum less than $250\mevc$, a cosine of the angle 
between the thrust axis of the event and the pion momentum less than 0.8 and a distance of closest 
approach to the primary vertex less than $0.5~\cm$. These criteria preferentially select pions from $\Dstar$ decays
which are abundant in \BB events.
\end{thebibliography}
\end{document}